# *Electrical Transport through a mechanically gated molecular wire*

C. Toher[1], R. Temirov[2,3], A. Greuling[4], F. Pump[1], M. Kaczmarski[4], M. Rohlfing[4], G. Cuniberti[1,5], F.S. Tautz[2,3] *

[1]Institute of Materials Science and Max Bergmann Center of Biomaterials, Dresden University of Technology, 01062 Dresden, Germany

[2]Institute of Bio-and Nanosystems (IBN 3), Forschungszentrum Jülich, 52425 Jülich, Germany

[3]JARA— Fundamentals of Future Information Technology, 52425 Jülich, Germany

[4] Department of Physics, University of Osnabrück, 49069 Osnabrück, Germany

[5]National Center for Nanomaterials Technology, POSTECH, Pohang 790-784, Republic of Korea

**Abstract**

A surface-adsorbed molecule is contacted with the tip of a scanning tunneling microscope (STM) at a pre-defined atom. On tip retraction, the molecule is peeled off the surface. During this experiment, a two-dimensional differential conductance map is measured on the plane spanned by the bias voltage and the tip-surface distance. The conductance map demonstrates that tip retraction leads to mechanical gating of the molecular wire in the STM junction. The experiments are compared with a detailed *ab initio* simulation. We find that density functional theory (DFT) in the local density approximation (LDA) describes the tip-molecule contact formation and the geometry of the molecular junction throughout the peeling process with predictive power. However, a DFT-LDA-based transport simulation following the non-equilibrium Green's functions (NEGF) formalism fails to describe the behavior of the differential conductance as found in experiment. Further analysis reveals that this failure is due to the mean-field description of electron correlation in the local density approximation. The results presented here are expected to be of general validity and show that, for a wide range of common wire configurations, simulations which go beyond the mean-field level are required to accurately describe current conduction through molecules. Finally, the results of the present study illustrate that well-controlled experiments and concurrent *ab initio* transport simulations that systematically sample a large configuration space of molecule-electrode couplings allow the unambiguous identification of correlation signatures in experiment.

* <u>Corresponding author:</u>
Prof. Dr. Stefan Tautz
Institut für Bio-und Nanosysteme 3
Forschungszentrum Jülich
52425 Jülich
Germany

Tel. +49 2461 61 4561
Fax: +49 2461 61 3907
e-mail: s.tautz@fz-juelich.de

1. Introduction

Any progress towards the vision of molecular electronics requires a thorough understanding of current conduction through single molecules. For this reason, an intense effort has been focused recently on transport experiments in metal/molecule/metal-junctions [1-9], utilizing mechanically controlled break junctions, nanofabricated junctions or the junctions of a scanning tunneling microscope (STM). Unfortunately, most single molecule transport experiments to date do not allow the structural characterization of the junction independent of its transport properties [10-12]. In this respect, STM-based approaches offer advantages. For example, in a recent publication, we have shown that a surface-adsorbed molecule can be chemically bonded to an STM tip at a defined position within the molecule and peeled off the surface by tip retraction [13], as shown schematically in Fig. 1. The tip-molecule bond forms spontaneously as soon as the tip has approached close enough to the specific bonding group in the molecule. Spectroscopic data recorded during tip retraction revealed a mechanical gating effect, in the sense that one of the molecular orbitals responds to the structural change and shifts with respect to the Fermi level of the substrate [13]. The experiments have been carried out on 3,4,9,10-perylene-tetracarboxylic-dianhydride (PTCDA) adsorbed at the Ag(111) surface. The system PTCDA/Ag(111) is a highly ordered and well-characterized organic-metal interface [14-27]. Molecules adsorb in flat orientation and form large, highly ordered and commensurate islands in the so-called herringbone structure. For details, we refer to ref. [14] and references therein.

In the present paper, we present more comprehensive data of this experiment and compare the experimental data to an *ab initio* study of transport through this molecular wire. To describe the conduction through a molecule theoretically from first principles, three demanding quantum mechanical problems have to be addressed [28], namely the geometric and electronic structures of the junction and the non-equilibrium current between two biased reservoirs across the molecular bridge. For the first two, we employ density functional theory (DFT), while the transport problem is solved by the non-equilibrium Green's function formalism (NEGF). Our analysis starts with a simulation of the contacting process during tip approach and the molecule-substrate bond cleaving process during tip retraction. This yields structural and energetic data which can be compared to experimental results. In the second step, the transport calculations are carried out for the experimentally validated geometries and compared to experimental transport data. In this way, we can separately compare structure and transport simulations to experiment. Moreover, the comparison is not limited to a single point in the configuration space of the molecular junction but includes a wide range of systematically and controllably varied configurations for each of which experimental transport data exist. The experiments thus provide a critical test of the theoretical methodology by effectively eliminating the possibility of fortuitous agreements between theory and experiments.

As we will see below, DFT fully confirms structural aspects of the contacting and bond cleaving scenarios deduced from experiment, while the transport simulations exhibit several differences compared to experiment. We find that these can be understood as a failure of the DFT/NEGF platform with the local density approximation (LDA) to account for the dynamical nature of electron-electron correlations in the transport orbital of the molecule.

The paper is organized as follows: In section 2, the most important details of the experimental technique are summarized. In section 3 we then report the results of the experiment, before turning to a detailed description of our simulation methods in section 4. Section 5 presents the simulation results (junction structure and energetics, electronic structure and transport). The comparison of the transport simulation to the corresponding experimental data is presented in section 6. The paper closes with conclusions and an outlook in section 7.

## 2. Experimental Methods

The experiments are carried out at 5K in a CREATEC low-temperature scanning tunneling microscope on the PTCDA/Ag(111) system. The Ag(111) surface was prepared by cycles of ion bombardment (0.8 keV Ar+ ions) and annealing at 550°C. PTCDA was deposited from a self-built evaporator onto the Ag(111) surface at room temperature.

In our experiments we use electrochemically etched tungsten tips. After standard tip cleaning procedures in vacuum and before starting actual experiments, the tip is dipped repeatedly into the Ag surface, until its density of states appears featureless; this is controlled by recording spectra of the Ag(111) surface state. Most likely, the tip apex is therefore a silver cluster. This conjecture is supported by comparison of our data to theory (cf. section 5.1).

For the comparison with the simulation, a careful calibration of the absolute z-scale in the experiment is essential. The calibration of the z-scale in our experiments proceeds as follows: (1) The prepared tip is stabilized above the centre of a PTCDA molecule at a set point with a current of $I$= 0.1 nA at a bias voltage of $V_b$=   0.34 V. (2) The tip is moved into contact with the molecule, discernable by a sudden deviation from the exponential approach curve that is characteristic of vacuum tunneling. We find that contact appears after approaching by $\Delta z = 4.32$ Å from the set point. (3) The z-coordinate of the tip at the set point is calculated by
$$z_{\text{set point}} = h_{\text{PTCDA}} + r_{\text{vdW}}^{\text{carbon}} + r_{\text{vdW}}^{\text{silver}} + \Delta z.$$
Here $h_{\text{PTCDA}} = 2.86$ Å is the height of the plane of the PTCDA molecule above the outermost lattice plane of the Ag(111) surface [16], and $r_{\text{vdW}}^{\text{carbon}} = 1.70$ Å and $r_{\text{vdW}}^{\text{silver}} = 1.72$ Å are the van der Waals radii of carbon and silver atoms, respectively [29]. From the above equation we find that the vertical distance between the centre of the tip apex atom at the set point and the outermost lattice plane of the Ag(111) surface is $z_{\text{set point}} = 10.6$ Å. (4) All experimental tip-surface distances $d_{\text{ts}}$ are referenced to $z_{\text{set point}}$, using the known and calibrated z-piezo constants of our microscope.

Experimental transport data in this paper are based on two-dimensional maps in which *dI/dV* is recorded nearly continuously as a function of bias voltage (step 1 meV) and tip-surface distance (step 1.5×10$^{-4}$ Å). The maps are recorded by rapid approach-retraction cycles of the tip (time for one cycle approx. 2.4s), i.e. the fast scanning variable in the ($V_b$,$d_{ts}$)-plane is the tip-surface distance $d_{ts}$. The closest point of the approach-retraction cycle is chosen such that at this distance the tip-molecule bond forms spontaneously. In our original experiment [13] we contacted the molecule once and then retracted the tip in steps, recording a full $dI/dV_b(V_b)$ spectrum after each retraction step. This technique has a disadvantage: If the contact breaks (e.g. due to high current through the molecular wire and associated inelastic processes), the tip must be brought back to the initial value of $d_{ts}$ at which the oxygen contact forms spontaneously and a new series of spectra must be started. In contrast, in the method employed in this paper we measure cuts through the ($V_b$,$d_{ts}$)-plane at constant $V_b$. After each cut the tip is approached to the sample again, giving it the chance to form a new contact with the molecule if the contact has been ruptured during retraction. Therefore, this way of doing the experiment is less vulnerable to tip-molecule ruptures. Using the new protocol it is possible to record with the STM differential conductance maps in the ($V_b$,$d_{ts}$)-plane that are dense, i.e. with an arbitrarily small step in $V_b$. These maps are comparable in quality to ($V_b$,$V_g$)-diagrams which are recorded in three terminal devices based on nano-junctions.

## 3. Experimental Results

When the STM tip approaches the molecule above the carboxylic oxygen, the oxygen atom spontaneously flips into contact with the tip apex [13]. This is discernible in the experiment by a sharp increase in the differential conductance. The most likely tip-surface distance at which this bond is formed is 6.65 Å (cf. the histogram in Fig. 3b). The formation of the chemical contact between tip and PTCDA molecule occurs exclusively at the carboxylic oxygen atoms. Through the flip into contact, a molecular wire which is bonded chemically to two electrodes is established (Fig. 1). The bond to the tip is locally confined to a single oxygen atom, while the bond to the surface is extended and involves several components, including the chemical interaction of delocalized π-electrons with metal *sp*-states, a sizable van der Waals contribution, and the remaining three local oxygen-metal bonds [14]. The tip-oxygen bond is strong enough to survive tip retractions during which the molecule is gradually cleaved from the surface, up to the point where the molecule is removed from the surface [13].

A typical differential conductance map as a function of tip-surface distance $d_{ts}$ and tip-sample bias voltage $V_b$ is displayed in Fig. 2a. It shows that upon stretching the Ag-tip/PTCDA/Ag(111) junction up to a tip-surface distance of $d_{ts} \approx 8$ Å the conductance resonance labeled A shifts from negative bias voltages toward zero. For distances larger than 8.2 Å, a sharp feature B that does not shift upon further tip retraction is observed at the Fermi level. Importantly, the evolution from A to B exhibits a sharp turn at the point where A and B touch. The spectra in Fig. 2b, generated from vertical cuts through Fig. 2a, indicate the coexistence of A and B in the range $d_{ts} \approx 8.0 \dots 8.2$ Å.

The $dI/dV_b(V_b)$ spectra presented in Fig.2b have been constructed by cutting through the ($V_b$,$d_{ts}$)-conductance map of Fig. 2a at constant $d_{ts}$. Hence, each data point in these $dI/dV_b(V_b)$ spectra originates from an individual tip-oxygen contact, but in their entirety these data points nevertheless form a smooth conductance spectrum that is characteristic of a specific tip-surface distance. This demonstrates the excellent reproducibility of the measured conductance for a given pair of variables ($V_b$,$d_{ts}$) in our experiments, as long as the tip geometry does not change. We note that the bias window in Fig. 2a and 2b is limited to approximately ± 100 mV by current-induced ruptures of the tip-molecule contact.

In a few cases we observe junctions which cannot be stretched beyond $d_{ts} \approx 8$ Å, where the contact always ruptures. These junctions have slightly different transport properties from the one shown in Fig. 2. Together with their different mechanical stability, the different transport properties indicate a different bonding geometry of the carboxylic oxygen atom to the tip. The different bonding may either originate from slight variations in the lateral position of the approach, with the result that the oxygen atom bonds to different positions on the tip, or from the tip structure itself. However, we stress again that over the time scale of 8 minutes, which are required to record a single conductance map, both the lateral position of the tip and its structure are stable, such that within one map the atomic structure of the tip-oxygen bond is always preserved.

## 4. Simulation Methods

### 4.1. General approach

Our *ab initio* simulations address geometric structure, electronic properties and transport of the molecular wire junction. For the first two, the density functional theory (DFT) package SIESTA is employed [30, 31]. We used the local density approximation (LDA), because this approximation yields more realistic structural parameters and adsorption energies for the PTCDA/Ag(111) interface than the generalized gradient approximation (GGA) [22]. Furthermore, the PTCDA/Ag(111) adsorption energetics resulting from LDA is closer to reality than that of other density functionals, as was shown in a recent RPA (random phase approximation) study [25]. The DFT equilibrium geometries and single-electron states serve as input for transport simulations based on the Landauer scattering approach (implemented using the non-equilibrium Green's function (NEGF) formalism) in which the scattering of transport electrons in the molecular bridge and the latter's electronic structure are solved self-consistently under applied bias [28, 32]. The NEGF calculations employ SMEAGOL [33], an extension to SIESTA. To cross-check, we have also used the gDFTB transport code [34], which gives similar results as SMEAGOL.

### 4.2. DFT simulations

All SIESTA simulations have been carried out on a separate molecule in the aligned orientation (molecule in bridge position on the Ag(111) surface, long axis parallel to the close-packed rows of Ag atoms) [22]. A local basis set with two s-, two p-, and one d-orbital for carbon as well as for oxygen, two s- and one p-orbital for hydrogen, and two s-, one p- and two d-orbitals for silver was employed. A small contraction energy of 0.002Ry was used [30, 31], yielding a spatial truncation of 3.57 Å, 2.81 Å, 3.45 Å and 4.54 Å for C, O, H, and Ag, respectively.

For the structural data (Figs. 3 to 5), the Ag-tip was modeled by a 10-atom tip tetrahedron, with 1, 3, 6 atoms in the first, second and third *fcc* layers, respectively (1, 4, 5 atoms in the first, second and third *bcc* layers for the W-tip), while the sample was modeled by three layers of silver atoms.

In order to simulate the molecule-substrate bond cleaving process on tip retraction, the tip was bonded to one of the carboxylic oxygen atoms (the equilibrium bonding distance in the simulation is $d_{\text{t-ox}}$ =2.16 Å for Ag-tips) and then removed on an arc in steps of 0.2 Å, each time followed by a full relaxation of all atomic coordinates of the molecule until residual forces fell below 0.01 eV/Å or 16 pN. If in the simulations the tip is retracted in a straight vertical line, the tip-oxygen bond breaks and the molecule cannot be peeled off the surface, because it would have to slide along the surface during peeling, as well. Such sliding is, however, too sensitive to numerical details to be described correctly by our present calculations. Rather than forcing the molecule to slide along the surface, we employ the opposite lateral movement of the tip during retraction, resulting in the aforementioned arc movement.

To generate the potential energy curves $E\left(d_{\text{s-ox}} ; d_{\text{ts}}\right)$ shown in Figs. 3 and 4, the molecule was relaxed for a series of constrained oxygen-surface distances $d_{\text{s-ox}}$, as above. These relaxed configurations were then inserted into tip-surface junctions of defined distances $d_{ts} = d_{\text{t-ox}} + d_{\text{s-ox}}$ and the total energy was calculated.

For the calculation of the partial density of states (PDOS) of the LUMO in DFT-LDA (as discussed in section 5.3 below), a supercell with 6 layers of silver atoms representing the surface and a 10-atom Ag tetrahedron attached to the backside of the substrate slab, representing the tip, was used. The PDOS was calculated by summing over the projections of all states of the junction onto the LUMO orbital of an isolated PTCDA molecule.

*4.3. NEGF simulations*

The self-consistent electronic structure and transport calculations have been carried out using the SMEAGOL NEGF code [33], which is built around the SIESTA DFT code. The SIESTA parameters were as described above for the structural calculations. The positions of the atoms in the PTCDA molecule were taken from the results of the structural calculations. As in the structural calculations, the molecule was placed on an Ag(111) substrate, and contacted by an Ag STM tip. However, in the transport calculations, the substrate was represented by 5 layers of silver to which was then connected a semi-infinite electrode, in the form of a self-energy added to the Hamiltonian prior to the calculation of the Green's function [28, 32, 33], while the tip in the transport calculations consisted of a 4 atom tetrahedron, with 1 and 3 atoms in the first and second *fcc* layers respectively, connected to a 3 layer thick silver surface, which was once again connected to a semi-infinite electrode. These alterations to the geometric structure of the junction were necessary for the transport calculations in order to screen the molecule, as the charge density of the system should be converged to the bulk value at the interface to the semi-infinite electrodes [33].

The complex part of the integral of the lesser Green's function leading to the charge density was evaluated using 50 energy points on the complex semi-circle, 38 points along the line parallel to the real axis and 20 poles. The integral over real energy necessary at finite bias is evaluated using at least 128 points. All of the transport calculations were carried out with periodic boundary conditions in the directions perpendicular to that of the transport and a mesh of 4 *k*-points.

In our calculations, the bias is applied symmetrically to the substrate and the tip, with the electrochemical potentials of each being shifted by $\pm\frac{eV}{2}$, whereas in the experiments the full bias is applied asymmetrically to just the substrate, i.e. the tip electrode is grounded. Additionally, there has been some discussion in the literature about where exactly the potential drops in a molecular junction, with the result depending on the specific properties of the device such as the screening length of the molecule[35, 36]. In SMEAGOL the potential and hence the voltage drop is calculated self-consistently via the Poisson equation, with the electronic structure of the junction also being calculated self-consistently incorporating the effect of the applied potential and the self-energies of the electrodes [33]. Therefore, the same physics results as for asymmetric bias, and the location of the potential drop should be accurately described, with any quantitative error being due to inherent shortcomings in DFT-LDA rather than the transport model being used.

## 5. Simulation Results

### 5.1. Junction structure: Contact formation

Fig. 3 displays calculated potential energies of the molecule in the junction (DFT-LDA), plotted against the oxygen-surface distance $d_{s\text{-ox}}$, for three different fixed tip-surface distances $d_{ts}$ as the tip approaches the surface (Fig. 3a and 3b) and as it is retracted again (Fig. 3c). According to the simulation, a spontaneous upward flip of the oxygen by δ ≈ 1.4 Å is expected at $d_{ts}$ ≈ 6.2 Å (Fig. 3b), because at this tip-surface distance the potential changes from a double-well to a single-well potential, characterized by a minimum below the tip and a saddle point at the former position of the oxygen atom above the surface [16]. Note that it is not only the oxygen atom which flips up. Due to the softness of the molecule, neighboring atoms follow the oxygen as it is attracted by the tip (Fig. 3b, inset). Fig. 3b shows the histogram of tip-surface distances at which in experiment the carboxylic oxygen atom flips into contact with the tip; the most likely distance is 6.65 Å [13]. The good agreement between the simulation and experiment (Δ$d_{ts}$ < 0.5 Å) confirms the interpretation of the experimentally observed conductance step [13] as originating from the flip into contact of the oxygen atom. The energy gain due to the flip is 0.6 eV, thus verifying the chemical nature of the tip-oxygen bond.

Because we use tungsten wire for preparing our tips, we have also calculated the potential energy profiles for a tip tetrahedron made from W atoms. The result is shown in Fig. 4. The flip into contact (with an amplitude of δ=2.44 Å) is predicted at a tip-surface distance $d_{ts}$ of 7.2 Å. Hence, for a W tip Δ$d_{ts}$ = 0.55 Å is thus slightly larger than for the Ag tip (Δ$d_{ts}$ = 0.45 Å), which, although not being conclusive in its own right, speaks in favor of the hypothesis that our tip apex atom is Ag rather than W. A second argument in favor of a Ag apex comes from experiment: On Au(111), with the tungsten tip prepared in exactly the same way as on Ag(111), we need to approach *closer* to the surface for the oxygen atom to flip into contact, although it is known that the bonding of the molecule to the Au(111) surface is *weaker* than to the Ag(111) surface. This observation shows that in the experiments on Au(111) the tip apex is less reactive than on Ag(111), which suggests that the tip is made of the respective substrate material, either Au or Ag depending on whichever is used as a substrate for tip preparation.

In a significant number of cases we observe in our experiments spontaneous tip-oxygen bond breaking on tip retraction, yielding hysteric bond-making and bond-breaking cycles. On the basis of the static potentials calculated in Fig. 3 and 4, this behavior cannot be rationalized. It must involve a hysteresis of potential landscapes which may be associated with distortions of the tip, occurring on bond formation and/or tip retraction.

*5.2. Junction structure: Bond cleaving*

Fig. 5 shows structural data related to the molecule-surface contact cleaving, calculated with DFT-LDA. Fig. 5a confirms the continuous nature of the peeling process, and in particular that the molecule is bent on tip retraction, since initially one end of the molecule (and hence also the centre of mass) is pulled up with almost no effect on the opposite end of the molecule (cf. animation in the supplement). Note that the lateral movement only sets in after ≈3 Å of tip retraction, in good agreement with the experimental finding that the wire conductance varies smoothly for the first ≈4 Å of tip retraction [13].

The continuous nature of the peeling process as a function of $d_{ts}$ is also reflected in the smooth behavior of the molecule-tip and molecule-surface bond energies as calculated in DFT-LDA (Fig. 5b). Although at small $d_{ts}$ the energy of the molecule-surface bond is significantly larger than that of the molecule-tip bond, the latter is strong enough to break the former. This results from the rather delocalized nature of the molecule-surface interaction, to which the atoms of PTCDA contribute about 0.1 eV each (to be overcome one after the other), while the molecule-tip bond with its energy of ≈1.3 eV is formed by one oxygen atom only. The detachment of the PTCDA from the tip that is observed in the simulation at $d_{ts} \approx 14$ Å (which is not found in experiment, where the molecule can be, but not always is, completely removed from the surface), is of no relevance in the present context, since it occurs outside the range of interest defined by the transport data in Fig. 2.

Summarizing the discussion of the structural DFT-LDA data in Figs. 3 to 5, we note that the simulation confirms (1) the formation of a molecular wire that interacts chemically with both contacts in the Ag-tip/PTCDA/Ag(111) STM junction, (2) the possibility to peel the molecule off the surface by tip manipulation, and (3) the smooth structural modification of the contact to the surface electrode in the process. In particular, we stress that the independently established z-scales of the experiment and the simulation coincide very well. We can thus conclude that the simulated structures of the molecular wire junction (as presented in the supplement) are an experimentally corroborated starting point for the electronic structure and transport simulations to which we now turn.

*5.3. Electronic structure of the junction*

In the vicinity of the Fermi level, the electronic structure of PTCDA/Ag(111) is dominated by the former lowest unoccupied molecular orbital (LUMO). On adsorption, the LUMO accepts electrons from the metal surface and shifts below the Fermi level [14, 22, 27]. In spite of its occupation, we refer to this state as the LUMO state throughout this paper.

The starting point of our simulation is an isolated (i.e. no molecular neighbors) and tip-contacted molecule on Ag(111), for which the peak of the PDOS of the LUMO state is predicted by DFT-LDA at -0.5 eV [37], compared to -0.45 eV in experiment [37]. The experimental value is estimated from ref. [13], where the LUMO of a PTCDA molecule in a herringbone layer contacted with the tip is found at -0.35 eV, and ref. [24], where it is shown that for uncontacted isolated molecules the LUMO shifts by approximately -0.1 eV with respect to molecules in a herringbone layer. We note, however, that the precise position of the resonance varies from experiment to experiment, presumably depending on the bonding geometry of oxygen atom on the tip. When the molecule is peeled off the surface, the junction changes its electronic structure. In DFT-LDA, the LUMO state shifts from -0.49 eV at $d_{ts}$= 5 Å to -0.03 eV at $d_{ts}$= 12 Å. At the same time, it becomes sharper, reducing its full width at half maximum from 0.27 eV (at $d_{ts}$= 5 Å) to 0.16 eV (at $d_{ts}$= 12 Å) [37].

*5.4. Transport through the junction*

The differential conductance spectrum is given by [28, 32]

$$\frac{dI}{dV_b}(V_b, d_{ts}) = \frac{2e}{h}\frac{d}{dV_b}\left(\int_{-eV_b/2}^{eV_b/2} T(E, V_b, d_{ts})dE\right),$$

and calculated in the NEGF formalism. $T(E, V_b, d_{ts})$ is the transmission of the wire at tip-surface distance $d_{ts}$ and applied bias voltage $V_b$ for incident electrons with the energy $E$.

In Fig. 6a, the zero bias transmission $T(E, 0, d_{ts})$ as calculated in NEGF is displayed for various $d_{ts}$. It is dominated by a single resonance which occurs for negative energies (i.e. below the Fermi energy). As for the PDOS of the LUMO state calculated in DFT-LDA, we observe a shift of this resonance toward the Fermi level and a sharpening as the molecule is peeled of the surface (increasing $d_{ts}$). We note that the peak in the zero bias transmission function in Fig. 6a tends to be slightly sharper than the DFT-LDA PDOS as calculated in ref. [37]. This discrepancy may arise from the differences in the way in which the tip and substrates are represented in the DFT-LDA and NEGF calculations, respectively (see section 4 above). These differences may affect the electronic structure of the junction, especially since the self-energies representing the semi-infinite electrodes dramatically change the boundary conditions for the wave functions in the junction, which in turn alters its density of states. However, if the PDOS of the LUMO is calculated within NEGF, it closely resembles the zero bias transmission $T(E, 0, d_{ts})$ as displayed in Fig. 6a. We can therefore identify the transmission resonance in $T(E, 0, d_{ts})$ with the LUMO state of PTCDA.

Fig. 7 shows examples of transmission functions at finite bias ($d_{ts}$ = 12 Å), calculated within NEGF. Both at positive and negative bias voltages, the transmission resonance nearly tracks the electrochemical potential of the surface electrode. Note, however, that for negative biases the transmission resonance is a little less strongly coupled to the Fermi level of the surface than for positive bias. The fact that the transmission function follows the electrochemical potential of the surface indicates an asymmetric coupling of the PTCDA molecule to tip and Ag(111) surface, in the sense that the coupling to the surface is much stronger. This is confirmed by Fig. 8 in which the potential drop across the tip/PTCDA/Ag(111) junction ($d_{ts}$ = 12 Å) is plotted. The figure indicates that most of the voltage drops between the tip apex atom and molecule as represented by its centre of mass, while the voltage drop between the molecule and the surface is small.

In Fig. 6b, calculated differential conductance spectra for various $d_{ts}$ are shown. For $V_b$ > -0.5 eV, $\frac{dI}{dV_b}(V_b, d_{ts})$ follows the zero bias transmission function $T(E, 0, d_{ts})$ closely, while for $V_b$ < -0.5 eV deviations become apparent. This similarity between $\frac{dI}{dV_b}(V_b, d_{ts})$ and the zero bias transmission function $T(E, 0, d_{ts})$ can be explained by the fact that energy difference between the centre of the transmission resonance and the electrochemical potential of the substrate remains approximately constant as the bias is applied, as described above. Because of this, the centre of this resonance enters the bias window, thus causing a maximum in the differential conductance, at a bias which is equivalent to the zero-bias energy difference between the centre of the resonance and the Fermi level. As a consequence, the differential conductance becomes approximately equal to the transmission,

$$\frac{dI}{dV_b}(V_b, d_{ts}) \approx \frac{2e^2}{h} T(eV_b, 0, d_{ts}).$$

Similar to the situation in tunneling spectroscopy, we can therefore interpret the differential conductance $\frac{dI}{dV_b}(V_b, d_{ts})$ as an indication of the equilibrium density of states, which in the present case is dominated by $\rho_{\text{LUMO}}(eV_b, d_{ts})$, the PDOS of the transport orbital. In the next section, the central result of the NEGF simulation as displayed in Fig. 6b will be compared to experiment. Therefore, we plot the peak positions of the $\frac{dI}{dV_b}(V_b, d_{ts})$ spectra from Fig. 6b alongside the experimental differential conductance map from Fig. 2a. The result is shown in Fig. 9a. Similarly, we compare the cut through the experimental conductance map in Fig. 2a at $V_b$=0 to the corresponding calculated quantity from Fig. 6b. The result is shown in Fig. 9b.

## 6. Discussion of the Transport Data

### 6.1. Reverse chemisorption

As the molecule is peeled off the surface, the peaks in all three quantities $\rho_{\text{LUMO}}(E, d_{ts})$ (ref. [37]), $T(E, 0, d_{ts})$ (Fig. 6a), and $\frac{dI}{dV_b}(V_b, d_{ts})$ (Fig. 6b) move toward the Fermi level or zero bias, respectively, and become sharper. The origin of this behavior can be understood as follows [13]: When the molecule is removed from the surface, its π-system dehybridizes from the surface. This leads to a destabilization of the charge transferred from the metal into the molecule. Since this charge mainly resides in the LUMO state, this orbital moves up towards lower binding energies. This is a direct effect of reducing the chemical interaction between molecule and surface. Therefore, we term the upward movement of the LUMO state "*reverse chemisorption*".

### 6.2. Comparison to experiment

In Fig. 9a the results of the transport simulations are compared to those of experimental data. Although the behavior is qualitatively similar, in that experiment also exhibits the shift of the LUMO state due to reverse chemisorption (in the range of $d_{ts}$= 7 to 8 Å), there is an important difference: A notable feature of reverse chemisorption in the NEGF simulation is the gradual reduction in the rate of energy shift as the LUMO state approaches the Fermi level. As a consequence, this means that the level reaches the Fermi energy only at $d_{ts} \gtrsim 12$ Å. In experiment, no such slowing down is discernable. On the contrary, between $d_{ts}$= 7 to 8 Å a rather steep slope of the peak position of 0.2 eV/Å is observed in experiment, and consequently the peak reaches the Fermi energy already at $d_{ts} \gtrsim 8$ Å, i.e. much earlier than in the simulation. This is also apparent Fig. 9b, where the zero bias conductance in the experiment reaches its peak ca. 4.5 Å earlier than in the simulation. Furthermore, in experiment there seems to be a clear distinction between the shifting (A) and the Fermi-level pinned (B) differential conductance peaks (Fig. 2), with a very small regime of coexistence, whereas the simulation predicts a more gradual evolution of the differential conductance which does not allow a distinction between resonances A and B to be made.

We have tested how sensitively the NEGF transport data react if the geometric junction structure is changed. Despite changing parameters such as the initial adsorption height of PTCDA, the tip trajectory and even the tip material, we did not manage to improve the agreement between transport experiment and theory substantially, although the end position of the LUMO state at large $d_{ts}$ varies slightly around the Fermi energy. This indicates that the disagreement between experiment and simulation is an essential consequence of the employed simulation method.

The end-point of the experiment is a PTCDA molecule that dangles off the tip and has lost contact to the Ag(111) surface. For this situation, NEGF predicts a transmission resonance that is 8 meV broad and occurs at +2 meV (cf. inset to Fig. 6a). Within the DFT-LDA-based NEGF formalism, this shows that (1) the FLUMO-tip hybridization is very small, presumably because the bond to the oxygen is localized at the periphery of the molecule, and also because the LUMO state has a π-orbital symmetry and so is aligned perpendicular to the tip axis in this configuration; and that (2) as a result of the chemical bond between the tip and the otherwise free molecule slightly less than one electron is transferred into the LUMO state of the molecule. In the NEGF simulation, the end point of the molecule-surface cleaving process is thus a molecule that is bonded *strongly* to the tip via local orbitals on the oxygen atom, but which at the same time has an unpaired electron in an essentially unbroadened LUMO state that is coupled *weakly* to the tip electrode. While this is a realistic description of the mechanical and electronic couplings between molecule and tip, it must be noted here that in such a situation electron correlation is expected to be so strong that its treatment as contained in NEGF and the underlying DFT-LDA is not adequate [37]. We will argue in section 6.4 that this inadequacy accounts for much of the deviation between experiment and the NEGF simulation in Fig. 9.

*6.3. Model calculation*

In order to elucidate the behavior of the DFT-LDA-based NEGF simulation, in particular the slowdown in the rate of energy shift of the LUMO state as it approaches the Fermi level, we have implemented a simple model calculation in order to rationalize the behavior of DFT-LDA in the simplest possible way. The molecular junction is modeled by a single energy level ε connected between two electron reservoirs. Due to the coupling to the reservoirs, this level is broadened so that it can be described by a Lorentzian density of states with width $\Gamma$.

The energy of the level has a mean-field type dependence on its occupation, given by the formula $\varepsilon(n) = \varepsilon_0 + nU$, where $n$ is the occupation of the level and $\varepsilon_0$ is the energy it would have at zero occupation [32, 38, 39]. The level is initially below the chemical potential of the reservoirs. To simulate the reverse chemisorption effect, the energy of the level is increased by adding a term $\beta$ to $\varepsilon_0$, such that $\varepsilon_0(\beta) = \varepsilon_0 + \beta$, while the coupling of the level to the electrodes, $\Gamma$, is simultaneously reduced so that $\Gamma(\beta) = \Gamma - 0.4\beta\Gamma$. The linear dependence of the level energy and its width on $\beta$ has been chosen for simplicity. The occupancy $n(\beta)$ of the level $\varepsilon$ follows from integrating its Lorentzian density of states up to the Fermi level, such that finally the level position of the level under reverse chemisorption, parametrized by $\beta$, is described by

$$\varepsilon\big(n(\beta)\big) = \varepsilon_0(\beta) + n(\beta)U.$$

The result of the model calculation for parameter values $\varepsilon_0(0) = -4.9\text{eV}$, $\Gamma(0) = 0.5\text{eV}$, and $U = 2\text{eV}$ is displayed in Fig. 10. Initially, the interacting level $\varepsilon\big(n(\beta)\big)$ follows the bare one, $\varepsilon_0(\beta)$, because variations in $n(\beta)$ are small. However, as the level is depopulated, its energy is reduced by the decreasing intra-orbital repulsion $n(\beta)U$. This tendency increasingly counteracts the upshift of $\varepsilon_0(\beta)$, with the result that, for the given parameters, $\varepsilon\big(n(\beta)\big)$ approaches the Fermi energy without crossing it.

The linear, mean-field behavior, which is represented by the term $n(\beta)U$ in the self-consistent level energy $\varepsilon\big(n(\beta)\big)$ in the above model, is a specific feature of LDA and various other approximations to the exchange-correlation functional that are used in DFT [40-42].

*6.4. Limitations of LDA as a mean-field theory*

The local density approximation (LDA) tends to work well for systems such as bulk metals in which the electronic states are highly delocalized, but it has problems with systems such as molecules which contain highly localized states where electronic correlation becomes more significant [43-45]. These problems manifest themselves in the form of the self-interaction error [42, 43, 45], the lack of the derivative discontinuity in the energy as a function of occupation [40, 41], and the failure to describe many-body phenomena such as the Kondo effect [46, 47].

One particular aspect where approximate exchange-correlation functionals deviate from the expected behavior of the true (but unknown) functional is in the treatment of fractional occupancy. In DFT, fractional charging of an open system (here: PTCDA molecule) coupled to a reservoir (here: electrodes) may be expressed as a statistical mixture of system ground states with different integer numbers of electrons [40]. It turns out that the total energy $E_{\text{tot}}$ has discontinuities in its derivative at integer values $Z - 1$, $Z$, $Z + 1$, ... of the total electron number $N = Z - 1 + n_{\text{LUMO}}$. It is a general result of DFT that the Kohn-Sham energy $\varepsilon_i$ of an orbital is given by $\varepsilon_i = \frac{\partial E_{\text{tot}}}{\partial n_i}$ [48]. Around integer occupancy of the LUMO state, i.e. $n_{\text{LUMO}} = 1$, one therefore gets two different Kohn Sham energies $\varepsilon_{\text{LUMO}}$, depending on whether the discontinuity is approached from above ($\varepsilon_{\text{LUMO}} = -A_z$) or below ($\varepsilon_{\text{LUMO}} = -I_z$), where $A_z$ and $I_z$ are the electron affinity and ionisation potential of the $Z$-electron system, respectively.

This derivative discontinuity can have a significant impact on electronic transport in molecular junctions [44]. Within LDA, which lacks this discontinuity, the Kohn-Sham eigenvalues exhibit a linear response to changes in their occupation, similar to the simple model used above, and thus show a "slow down" of the rate of energy shift as the LUMO state (and the corresponding transport resonance) approach the Fermi level. On the other hand, with the true exchange-correlation functional containing the derivative discontinuity, there is no dependence of the Kohn-Sham eigenvalue on $n$ as long as $n$ is non-integer. Therefore, in the exact DFT description of reverse chemisorption, $\varepsilon$ must follow $\varepsilon_0$ for $n_{\text{LUMO}} > 1$. A behavior like this is observed in experiment (Fig. 2a up to ≈ 8 Å). This suggests that the discrepancy between experiment on the one hand and the DFT-LDA-based NEGF simulation on the other can be accounted for by the incorrect mean-field treatment in LDA.

*6.5. Beyond mean-field theory*

For a conceptually correct description of transport through the tip/PTCDA/Ag(111) junction, we need to achieve an atomistic description of the junction that yields the correct electronic spectrum under non-equilibrium conditions. An atomistic description of the junction structure is, e.g., provided by DFT in the local density approximation. However, the calculation of a conceptually well-founded electronic spectrum demands that dynamical correlation beyond the mean-field level must be taken into account. This is difficult to achieve in DFT, which is a ground state theory, although it is true that the correct excitation properties are implicitly contained in the exact ground state electron density, since the latter uniquely defines the external potential that in turn fully defines the Hamiltonian. In particular, in determining the electronic spectrum we must look beyond the Kohn-Sham eigenvalues of DFT, because there is no guarantee that those will yield the correct and complete spectral density [41, 49], even if the exact exchange-correlation functional of DFT was known and employed.

In ref. [37], we have presented a calculation scheme for the electronic spectra of the tip/PTCDA/Ag(111) junction at various tip-surface distances. It combines DFT in the local density approximation for providing atomic details of the junction structure, many-body perturbation theory (MBPT) in a simplified GW approximation [50] for non-local correlation, and the numerical renormalization group approach (NRG) for dynamical correlation. The agreement between experimental conductance map in Fig. 2a and the simulated excitation spectra of ref. [37] is good. In particular, the calculation reveals that feature A in Fig. 6a is a transport resonance which is due to a single single-particle peak that approaches the Fermi level for increasing $d_{\text{ts}}$, while feature B is a many-body (Kondo) resonance that is part of a three-peak structure, with the two single-particle side bands appearing outside the bias voltage window that is accessible in experiment.

As a property of the excitation spectrum, the Kondo resonance is not contained in the spectrum of Kohn-Sham levels of DFT, even if the exact exchange-correlation functional was used, so it is not surprising that the DFT-LDA-based NEGF simulation presented here does not show this resonance. Note that in the scheme of ref. [37] the electronic spectrum is calculated under equilibrium conditions. The agreement with experimental data proves that the effects of non-equilibrium (i.e. finite bias) are small. According to the results of the present paper, this is to be expected, because the junction is electronically asymmetric and voltage drops mainly across the tip-molecule contact (cf. section 5.4).

## 7. Conclusion and outlook

In conclusion, we have presented a precise experiment/simulation comparison for the molecular junction Ag-tip/PTCDA/Ag(111). We find that our atomistic non-equilibrium transport calculation using a Hamiltonian based on DFT-LDA does predict the mechanical gating effect of reverse chemisorption, i.e. the shift of the LUMO resonance of the molecule as the junction is stretched and the molecule is peeled off the surface. However, because in electron correlation is treated only at the mean-field level in the DFT-LDA-based NEGF formalism, it fails to accurately reproduce the experimental results. The electron correlations manifest themselves in a Kondo resonance that is observed in the Ag-tip/PTCDA/Ag(111) junction, although the PTCDA molecule itself is intrinsically non-magnetic.

The results described in this paper thus shed light on the actual behavior of a molecular orbital as a function of parameters such as the coupling strength to the electrodes and applied bias, and also allows us to infer the behavior of the orbital as a function of its occupation. In particular, the high level of geometric control present in both the measurements and the calculations indicate precisely how standard exchange-correlation functional such as LDA and GGA differ from reality, which in turn will assist in efforts to develop new, more accurate functionals [51].

We stress that the importance of accurately describing electron correlation, exemplified here for the tip/PTCDA/Ag(111) junction, is of broad relevance for molecular wires and more generally π-conjugated molecular adsorbates, because, firstly, the chemical interaction between the molecule and the electrodes (or surface) in many cases leads to charge transfer, secondly, typical $U$ values in molecules are substantial (for PTCDA we have calculated 3.0 eV for the free molecule and 0.9 eV for the adsorbed molecule, where this reduction is mainly due to screening by the metal [37]), and thirdly the hybridization between states of the molecule and the metal is mostly not large enough to dilute the local nature of the transport orbital.

In present case it has turned out that taking proper account of electron correlation in the electronic spectrum is more important for the accurate description of transport through the metal/molecule/metal junction than including the effects of non-equilibrium. While the importance of correlation is universal for reasons given above, the relative un-importance of non-equilibrium effects is a special feature of the asymmetrically coupled junction studied here, where the main transport resonance follows the electrochemical potential of the substrate. To establish an approach to predicting transport through molecular wires that is generally applicable, the combination of the correct spectrum with non-equilibrium is required.

Beyond the field of transport, our findings also have implications for the study of the adsorption of large molecules on surfaces. In the quest to understand the surface bonding of complex adsorbates, attempts have been made to tune their chemical interaction with the surface by systematic modifications of the molecule [52]. In that approach, however, gradual changes of the bonding properties are impossible to achieve. In this work, we have presented a direct way to force the molecule along a certain reaction coordinate in the interaction space between the molecule and the surface. With this method, it is thus possible to study the continuous evolution of surface chemical bonds.

## Acknowledgments

Computational facilities were provided by the Zentrum für Informationsdienste und Hochleistungsrechnen in TU Dresden and by the Jülich Supercomputing Centre at the FZJ. We gratefully acknowledge the financial support of the Deutsche Forschungsgemeinschaft in the framework of the priority program SPP 1243, the South Korean Ministry of Education, Science, and Technology Program, Project WCU ITCE No. R31-2008-000-10100-0, and from ECEMP – European Center for Emerging Materials and Processes Dresden (Project A2).

## Supplementary material

gif-animation of bond cleaving process.

**Figure Captions**

**Fig. 1** Schematic of the STM-based single molecule transport experiment across the Ag-tip/PTCDA/Ag(111) junction. The STM tip contacts the PTCDA molecule at one of its carboxylic oxygen atoms. As the tip is retracted, the molecule is peeled off the surface. This breaks the π-interaction between the molecule and the surface, as indicated by the thick dashed line.

**Fig. 2** (color online) Experimental results. **(a)** Differential conductance map in the $V_b$-$d_{ts}$ plane. Grey scale quantifies conductance, ranging from 0 (black) to 0.10 $G_0$ (white). $dI/dV$ was detected with a lock-in amplifier, modulation amplitude 4 mV and modulation frequency 2.9 kHz. **(b)** $dI/dV(V_b)$ obtained as cuts through the map in panel (a), at tip-surface distances $d_{ts}$ indicated by blue and red lines in panel (a) (blue from bottom to top: $d_{ts}$ =7.86, 7.97, 8.05, 8.14, 8.23, 8.32, 8.41 Å; red from bottom to top: $d_{ts}$ = 9.67, 9.78, 9.89, 10.01, 10.12, 10.24, 10.35 Å). Spectra are vertically offset. Red spectra are magnified by a factor of 5 with respect to blue spectra. For each spectrum, the zero conductance level is visible at the left. The labels A and B mark two peaks in the conductance, cf. main text for details.

**Fig. 3** (color online) Tip-oxygen contact formation for an Ag tip. *Black dots:* Simulated potential energy profiles of a fully relaxed PTCDA molecule in the tip-surface junction as a function of the carboxylic oxygen-surface distance $d_{s\text{-}ox}$, at tip-surface distances **(a)** $d_{ts}$ = 9.0 Å **(b)** $d_{ts}$ = 6.2 Å, **(c)** $d_{ts}$ = 7.8 Å. In panel (a) (approaching tip before contact formation) and (c) (retracting tip after contact formation), the left minimum of the potential energy profile corresponds to the oxygen atom bonded to the surface (bonding distance $d_{s\text{-}ox}$=2.68 Å [16]), while the right minimum corresponds to the oxygen atom bonded to the Ag-tip (bonding distance $d_{t\text{-}ox}$=2.16 Å). The saddle point potential in panel (b) corresponds to the tip-sample distance at which the oxygen atom flips by 1.36 Å from the surface to the tip, as indicated by the *black arrow*. *Large grey circles* indicate silver atoms of surface (left) and tip apex (right). *Small pink circles* indicate carboxylic oxygen atoms. Atoms are drawn to scale, with covalent radii of 1.45 Å and 0.73 Å for Ag and O, respectively [29]. In panel (b), the experimental histogram for the tip positions at which the spontaneous flip into contact occurs is plotted for PTCDA molecules in the herringbone structure [13]. The maximum of the histogram is at 6.65 Å. The insets show corresponding junction geometries, with tip atoms in *grey*, carbon atoms in *green*, oxygen atoms in *red*, and hydrogen atoms in *white*.

**Fig. 4** (color online) Tip-oxygen contact formation for a W tip. *Black dots:* Simulated potential energy profiles of a fully relaxed PTCDA molecule in the tip-surface junction as a function of the carboxylic oxygen-surface distance $d_{s\text{-}ox}$, at tip-surface distances **(a)** $d_{ts}$ = 8.8 Å **(b)** $d_{ts}$ = 7.2 Å, **(c)** $d_{ts}$ = 8.0 Å. In panel (a) (approaching tip before contact formation) and (c) (retracting tip after contact formation), the left minimum of the potential energy profile corresponds to the oxygen atom bonded to the surface (bonding distance $d_{s\text{-}ox}$=2.68 Å [16]), while the right minimum corresponds to the oxygen atom bonded to the W-tip (bonding distance $d_{t\text{-}ox}$=2.08 Å). The saddle point potential in panel (b) corresponds to the tip-sample distance at which the oxygen atom flips by 2.44 Å from the surface to the tip, as indicated by the *black arrow*. *Large grey circles* indicate silver atoms of surface (left) and tip apex (right). *Small pink circles* indicates carboxylic oxygen atoms. Atoms are drawn to scale, with covalent radii of 1.46 Å and 0.73 Å for W and O, respectively [29]. In panel (b), the experimental histogram for the tip positions at which the spontaneous flip into contact occurs is plotted for PTCDA molecules in the herringbone structure [13]. The maximum of the histogram is at 6.65 Å. The insets show corresponding junction geometries, with tip atoms in *grey*, carbon atoms in *green*, oxygen atoms in *red*, and hydrogen atoms in *white*.

**Fig. 5** Structure and energetics of the molecule-surface cleaving process. **(a)** Simulated vertical displacement of the centre of mass of the molecule (*circles*), lateral (*triangles*) and vertical displacement (*squares*) of the dianhydride group on the opposite side of the molecule as a function of tip-surface distance $d_{ts}$, evaluated for relaxed junction geometries. **(b)** Simulated PTCDA-surface interaction (*dotted line*), PTCDA-tip interaction (*dashed line*), and PTCDA distortion energies (*full line*) as a function of tip-surface distance $d_{ts}$, evaluated for relaxed junction geometries. Vertical dotted line indicates the flip into contact from Fig. 3.

**Fig. 6** Simulated transport spectra, calculated in DFT-LDA-based NEGF. **(a)** Zero bias transmission $T(E, V_b=0, d_{ts})$ for various tip-surface distances $d_{ts}$ as shown on the right. Inset: $T(E, V_b=0)$ through a tip-suspended PTCDA molecule, i.e. a molecule that is contacted via one of its carboxylic O atoms to the Ag-tip, but disconnected from the surface (minimum distance > 5 Å). Energy zero is the Fermi level. **(b)** Differential conductance $dI/dV(V_b, d_{ts})$ in units of the conductance quantum $G_0$ as a function of tip-surface distance $d_{ts}$ as given on the right.

**Fig. 7** Simulated transmission spectra $T(E,V_b)$ at finite bias, calculated in DFT-LDA-based NEGF. Tip-surface distance of is $d_{ts}$ =12 Å. **(a)** Positive bias. **(b)** Negative bias. Chemical potentials in the two electrodes are indicated by vertical dotted lines. In the simulation, the bias is applied symmetrically.

**Fig. 8** Simulated potential profile across the junction, calculated in DFT-LDA-based NEGF. Bias voltages are ±0.2 eV for solid and dashed line, respectively. Tip-sample distance is $d_{ts}$ =12 Å.

**Fig. 9** (color online) Comparison of simulated transport spectra (DFT-LDA-based NEGF) with experimental data. **(a)** Simulated *dI/dV* peak positions (*black circles*) from Fig. 6b and experimental *dI/dV* map from Fig. 2a. **(b)** Differential conductance in units of the conductance quantum $G_0$ as a function of tip-surface distance $d_{ts}$. *Black circles*: simulated 4mV differential conductance calculated as $\frac{I(4\text{meV}, d_{ts}) - I(2\text{meV}, d_{ts})}{2\text{meV}}$. *Solid lines*: Zero bias horizontal cuts through 7 differential conductance maps of the type displayed in Fig. 2a, measured under identical conditions. The thick red line is a cut through Fig. 2a. The comparison between the various solid lines illustrate the degree of reproducibility of the experimental data in Fig. 2.

**Fig. 10** (color online) Self-consistent mean-field model calculation for the reverse chemisorption of a doubly degenerate level that mimics the behavior of LDA. The junction configuration for increasing tip-surface distance is plotted on the horizontal axis ("reverse chemisorption parameter $\beta$"). *Dashed blue line*: bare level $\varepsilon_0(\beta)$, shifting linearly due to reverse chemisorption. *Solid blue line*: self-consistent mean-field level position $\varepsilon(\beta) = \varepsilon_0(\beta) + n(\beta)U$. *Red dotted lines*: Kohn-Sham energies of the model system as expected in exact DFT. Once $\varepsilon_0(\beta)$ reaches the Fermi energy, whence $n_{\text{LUMO}} = 2 \to 1$, the Kohn-Sham level $\varepsilon(\beta)$ jumps discontinuously from $\varepsilon_0(\beta) = -A_z$ for $n_{\text{LUMO}} > 1$ to $-I_z$ for $n_{\text{FLUMO}} < 1$. In the mean-field model, both the level occupancy and the level position vary more smoothly. At the beginning and end of the blue solid line, level occupancies are $n_{\text{FLUMO}} = 1.9$ and $n_{\text{FLUMO}} = 1.3$, respectively. Further discussion in the main text.

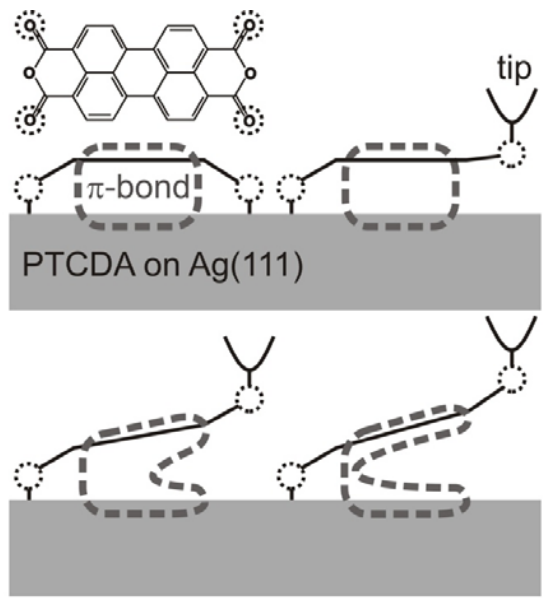

Fig. 1 Toher et al. 2010

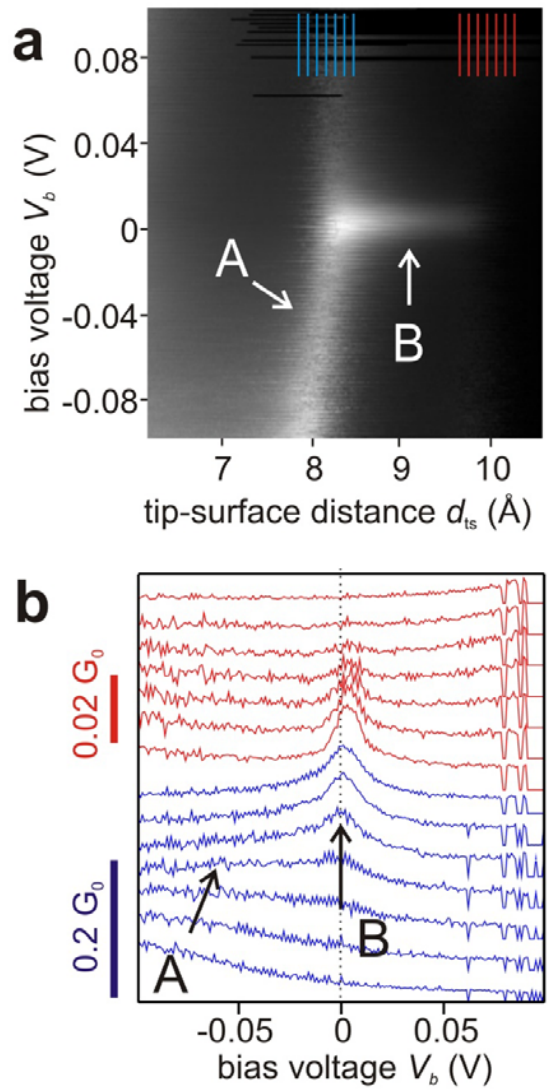

Fig. 2   Toher et al. 2010

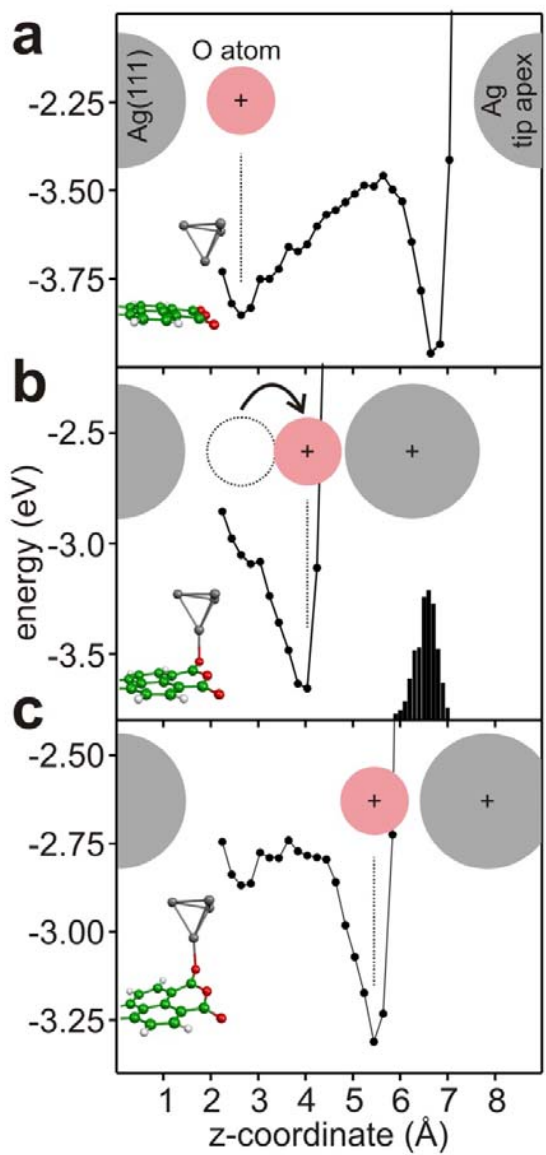

Fig. 3 Toher et al. 2010

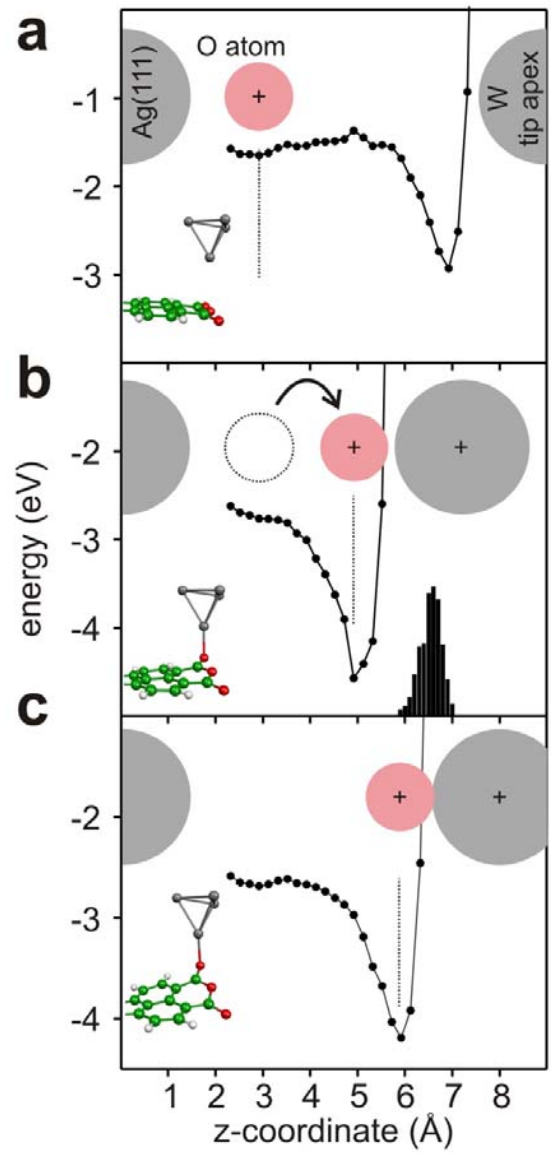

Fig. 4 Toher et al. 2010

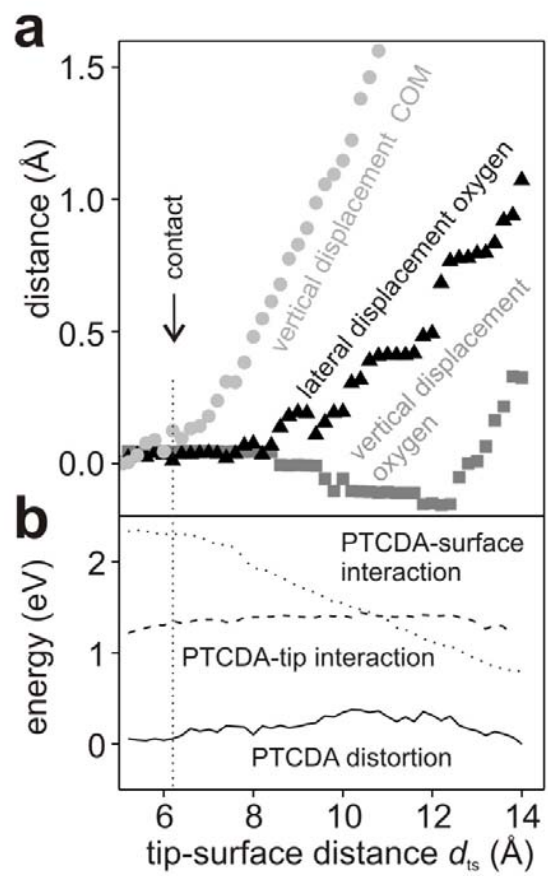

Fig. 5 Toher et al. 2010

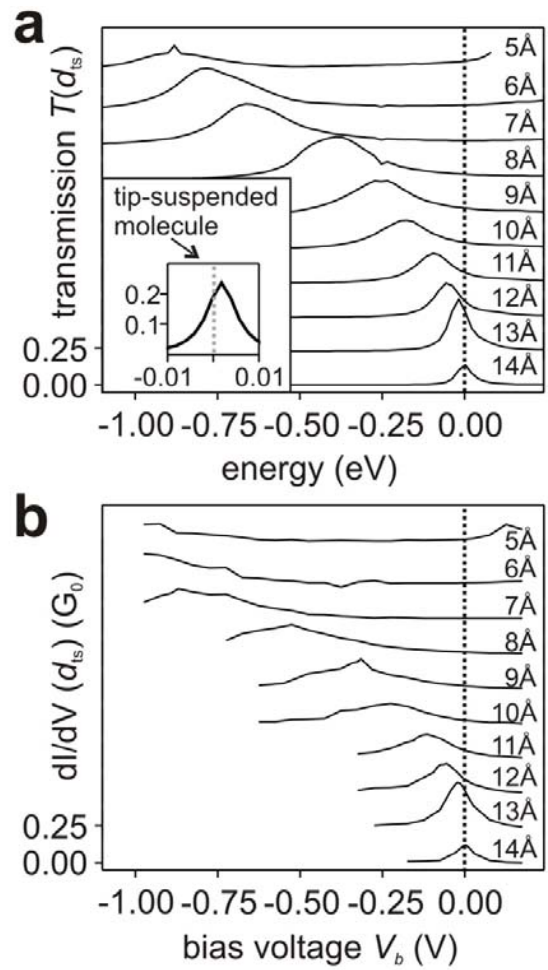

Fig. 6  Toher et al. 2010

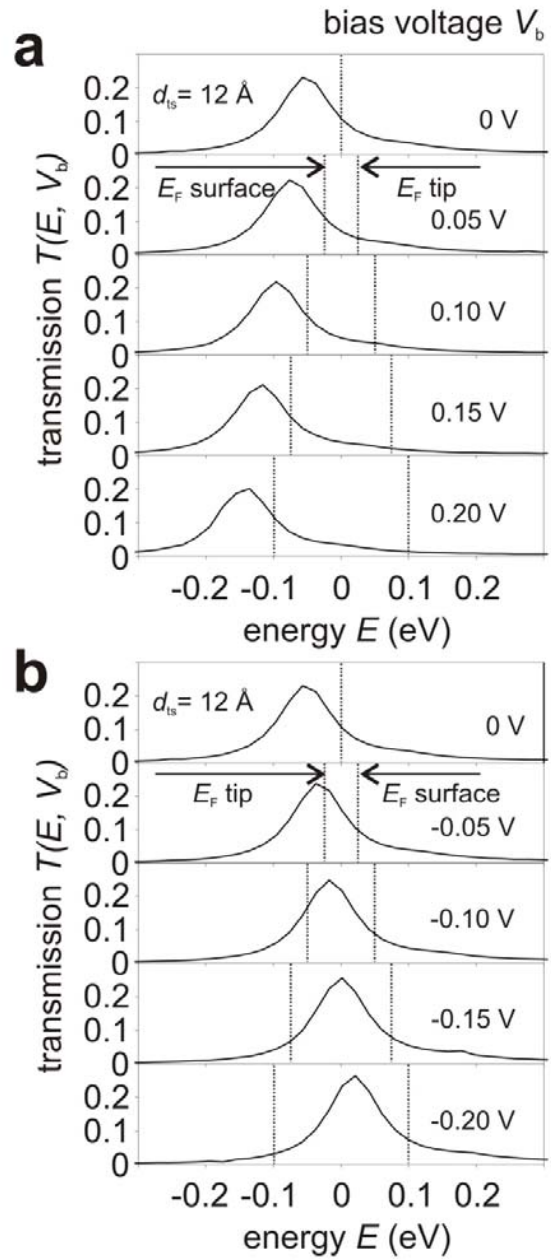

Fig. 7  Toher et al. 2010

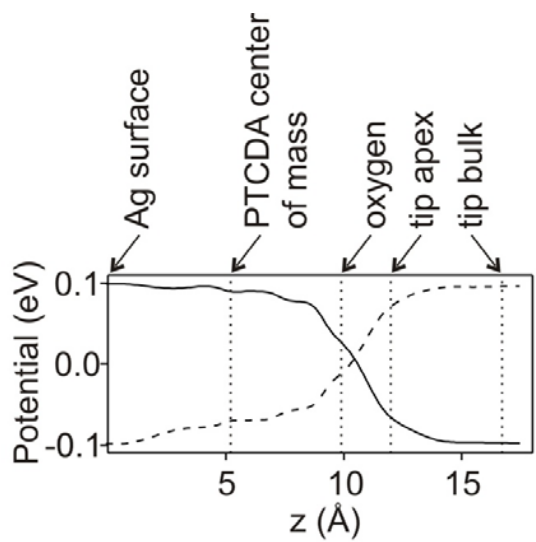

Fig. 8   Toher et al. 2010

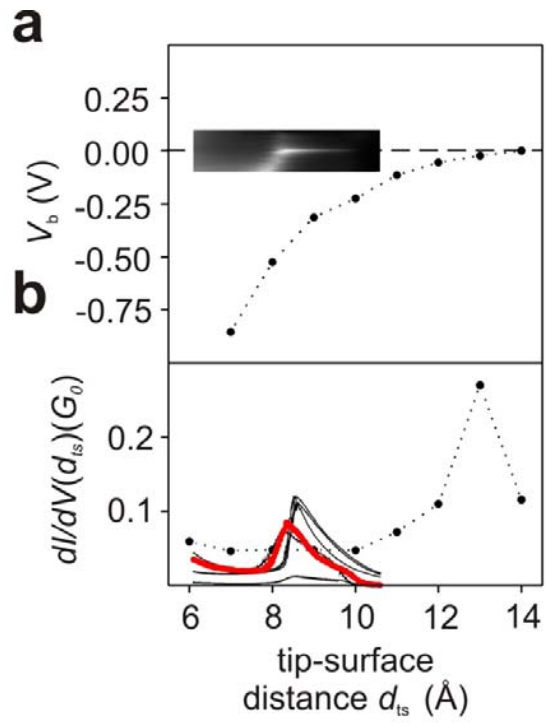

Fig. 9 Toher et al. 2010

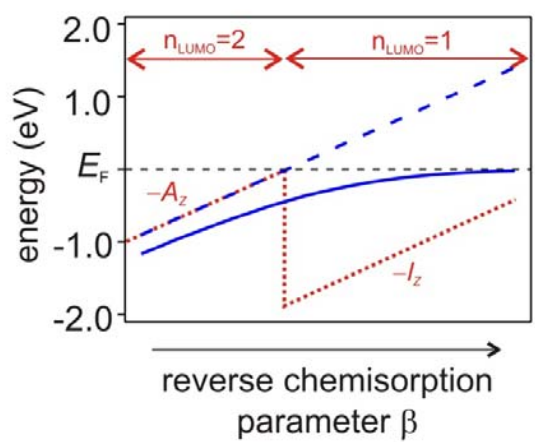

Fig. 10   Toher et al. 2010